\documentclass[preprintnumbers,prd,twocolumn,showpacs,floatfix,preprintnumbers,superscriptaddress,nofootinbib]{revtex4-2}

\usepackage{graphicx}
\usepackage{epsfig}
\usepackage{bm}
\usepackage{amssymb}
\usepackage{float}
\usepackage{amsmath}
\usepackage{subfigure}
\usepackage{dcolumn}
\usepackage[colorlinks]{hyperref}
\usepackage[usenames,dvipsnames]{color}
\hypersetup{
     breaklinks=true,
    pdfstartview={FitH},    
    colorlinks=true,       
    linkcolor=blue,          
    citecolor=red,        
    filecolor=magenta,      
    urlcolor=blue,           
    anchorcolor=green,      
    linktocpage=true
}
\usepackage{orcidlink}

\def\doi{http://doi.org}

\allowdisplaybreaks

\begin{document}

\title{Derivative hierarchy as the origin of kernel-dependent trends in Gaussian process reconstructions of the hubble parameter}

\author{Afaq Maqsood\orcidlink{0009-0000-3084-9169}}
\email{afaq.res@gmail.com}
\affiliation{Department of Physics, Jamia Millia Islamia, New Delhi, 110025, India}

\begin{abstract}
We perform a model-independent reconstruction of the cosmic expansion history using Gaussian Process regression, investigating how the smoothness properties of covariance kernels influence the inferred Hubble parameter. Using 32 cosmic chronometer measurements, we reconstruct $H(z)$ with the Matérn $3/2$, $5/2$, $7/2$, $9/2$, and Squared Exponential kernels, which form a well-defined hierarchy of differentiability. Across this hierarchy, we observe a systematic and monotonic trend in which increasing kernel smoothness is associated with progressively lower reconstructed values of the present-day Hubble constant $H_0$, accompanied by reduced statistical uncertainties. Rather than indicating a statistical preference for a specific kernel, this behaviour reflects the sensitivity of non-parametric reconstructions to the assumed smoothness prior encoded in the covariance function. The same ordering is reflected in descriptive goodness-of-fit statistics and remains stable when incorporating recent DESI DR2 BAO measurements and performing jackknife resampling tests, demonstrating robustness against individual data points and dataset variations. Additional analysis of derivative reconstructions shows that differences in kernel differentiability propagate into the local slope of the expansion history, providing a consistent interpretation of the observed ordering in $H_0$. Our results highlight that kernel smoothness plays an important role in Gaussian Process reconstructions and should be carefully accounted for when interpreting non-parametric cosmological inferences.\\

\vspace{2pt}
\noindent\textit{Keywords:} Cosmology, GPR, Machine learning

\end{abstract}

\maketitle
\flushbottom

\section{Introduction}
\label{sec:intro}

The accelerated expansion of the Universe, first revealed by observations of type~Ia supernovae~\cite{SupernovaCosmologyProject:1998vns,SupernovaSearchTeam:1998fmf}, has since been independently confirmed by a wide range of cosmological probes, including anisotropies in the cosmic microwave background~\cite{Planck:2013pxb,Planck:2018vyg} and measurements of baryon acoustic oscillations~\cite{DESI:2024kob,DESI:2025zgx,10.1093/mnras/,eBOSS:2018cab,2017MNRAS.470.2617A,2017A&A...608A.130D,DESI:2024mwx,DESI:2024uvr,DESI:2024lzq,DESI:2024aqx}. Within the standard $\Lambda$CDM framework, this phenomenon is attributed to a cosmological constant, which provides an excellent fit to a wide range of cosmological observations. 
Despite its empirical success, however, persistent challenges remain, most notably the Hubble tension---the discrepancy between local measurement \cite{Riess:2021jrx} and early-Universe \cite{Planck:2018vyg}determinations of the Hubble constant, $H_0$. 
This tension has prompted renewed scrutiny of both observational systematics and methodological assumptions that may influence cosmological inference.

Among the observational probes of the cosmic expansion history, cosmic chronometers (CC) 
\cite{Zhang_2014, Stern:2010cv,Moresco:2012jh,Moresco_2016,10.1093/mnras/stx301,10.1093/mnrasl/slv037} 
provide a particularly valuable and direct approach. 
By exploiting differential age measurements of passively evolving galaxies, CC yield model-independent estimates of the Hubble parameter $H(z)$ without relying on an assumed cosmological background. 
The current compilation of  CC-32 measurements spans the redshift range $0.07<z<1.965$, offering a clean dataset for reconstructing the expansion history while minimizing assumptions about the underlying cosmological model.

Gaussian Process (GP) regression has emerged as a powerful non-parametric framework for such reconstructions \cite{Rasmussen:2005gp,Holsclaw:2010sk,Holsclaw:2010nb,Holsclaw:2011wi}, allowing cosmological functions and their derivatives to be inferred directly from data without imposing a predefined functional form. 
GP methods have been widely applied to reconstruct the Hubble parameter, the dark energy equation of state, and related quantities\cite{Busti:2014aoa,Busti:2014dua,Seikel:2012uu,Seikel:2012cs,Seikel:2013fda,Shafieloo:2012ht,Mukherjee:2022yyq,Mukherjee:2022lkt,Mukherjee:2020vkx,Ruchika:2025mkx,Velazquez:2024aya,Mukherjee:2024ryz,Favale:2023lnp,Jiang:2025ilh,Favale:2025mgk,Jiang:2024xnu,Zhang:2025bmk}. 
However, an essential aspect of GP methodology that has received comparatively little attention in cosmological applications is the choice of covariance kernel, which encodes prior assumptions about smoothness, differentiability, and correlation structure.

The kernel choice in GP regression acts as a prior on the reconstructed function and can therefore significantly influence inferred cosmological quantities\cite{Mukherjee:2022yyq,Mukherjee:2022lkt,Mukherjee:2020vkx,Ruchika:2025mkx,Velazquez:2024aya,Mukherjee:2024ryz,Favale:2023lnp,Johnson:2025blf}. 
In particular, the Matérn family of kernels introduces a well-defined derivative hierarchy: a Matérn kernel with smoothness parameter $\nu$ admits reconstruction of derivatives only up to order $\lfloor \nu \rfloor$\cite{Mukherjee:2024ryz,Favale:2023lnp}. 
This hierarchy naturally interpolates between the relatively flexible Matérn $3/2$ kernel and the infinitely differentiable Squared Exponential kernel, providing a systematic framework for assessing how increasing smoothness assumptions affect cosmological reconstructions.

In this work, we investigate how kernel selection within this derivative hierarchy impacts the reconstruction of the Hubble parameter from cosmic chronometer data. 
Focusing primarily on the CC-32 measurements, we analyze reconstructions obtained using the Matérn $3/2$, $5/2$, $7/2$, and $9/2$ kernels, as well as the Squared Exponential kernel, thereby isolating the role of kernel-induced smoothness without additional complications from multi-probe combinations. 
This controlled setup allows us to directly compare reconstructed expansion histories, hyperparameters, and inferred values of $H_0$ across kernels. 
In addition, we examine how these kernel-dependent trends persist when recent six
BAO measurements of $D_H/r_d$ \cite{DESI:2024kob,DESI:2025zgx,10.1093/mnras/,eBOSS:2018cab,2017MNRAS.470.2617A,2017A&A...608A.130D,DESI:2024mwx,DESI:2024uvr,DESI:2024lzq,DESI:2024aqx}
are incorporated through a statistical consistency analysis of the reconstruction.

We find that kernel choice constitutes a non-negligible source of systematic uncertainty in GP-based cosmological inference. 
Progressively smoother kernels yield systematically lower values of $H_0$ with reduced uncertainties, accompanied by shorter inferred correlation lengths, while more flexible kernels provide better fits to local CC features. 
This behaviour is reflected in a clear hierarchy of $\chi^2$ values that remains robust under the inclusion of DESI BAO data\cite{DESI:2025zgx}. 
To further assess the stability of our conclusions, we perform a jackknife resampling of the CC-32 dataset, finding that neither the kernel ordering nor the inferred $H_0$ trends are driven by individual data points. These results demonstrate that kernel selection is not merely a technical detail but a physically relevant methodological choice, with direct implications for non-parametric reconstructions of the cosmic expansion history and for interpretations of current $H_0$ measurements.

\section{Gaussian processes}

Gaussian Processes (GPs) provide a nonparametric Bayesian framework for reconstructing an unknown function directly from data \cite{Rasmussen:2005gp,Holsclaw:2010sk,Holsclaw:2010nb,Holsclaw:2011wi}. A GP is specified by a mean function $\mu(x)$ and a covariance kernel $k(x,x')$,
\begin{equation}
f(x) \sim GP(\mu(x), k(x,x')) ,
\end{equation}
with
\begin{equation}
\mu(x)=E[f(x)], \qquad 
k(x,x') = E[(f(x)-\mu)(f(x')-\mu)] .
\end{equation}

For input points $X=\{x_i\}$, the covariance matrix is
\begin{equation}
[K(X,X)]_{ij} = k(x_i,x_j).
\end{equation}
Given data $y$ with noise covariance $C$, the joint distribution of observed data and predictions at test points $\hat X$ is
\begin{equation}
\begin{pmatrix}
y \\ \hat f
\end{pmatrix}
\sim \mathcal N\!\left(
\begin{bmatrix}
\mu \\ \hat\mu
\end{bmatrix},
\begin{bmatrix}
K(X,X)+C & K(X,\hat X) \\
K(\hat X,X) & K(\hat X,\hat X)
\end{bmatrix}
\right).
\end{equation}

The GP predictive distribution is
$\hat f \sim GP(a,A)$,
with mean and covariance
\begin{align*}
a &= \hat\mu + K(X,\hat X)^{T} [K(X,X)+C]^{-1} (y-\mu), \\
A &= K(\hat X,\hat X) 
    - K(X,\hat X)^{T} [K(X,X)+C]^{-1} K(X,\hat X).
\end{align*}

Hyperparameters of the kernel are fixed by maximizing the log marginal likelihood
\begin{equation}
\ln P(y)
 = -\frac12 (y-\mu)^{T} [K+C]^{-1} (y-\mu)
   -\frac12 \ln |K+C|
   -\frac{p}{2} \ln (2\pi).
\end{equation}
In the present analysis we adopt a zero mean function,
$\mu(z)=0$ \cite{Mukherjee:2022yyq,Mukherjee:2022lkt,Mukherjee:2020vkx,Ruchika:2025mkx,Johnson:2025blf}, which is a standard choice in cosmological Gaussian
Process reconstructions as it minimizes prior assumptions
and allows the expansion history to be reconstructed primarily from the observational data.

We have nevertheless verified that adopting alternative simple mean functions, such as a constant mean, produces statistically consistent reconstructions of $H(z)$. 
Although small shifts in the inferred value of $H_0$ may occur, the characteristic hierarchy associated with the covariance kernel smoothness remains preserved, indicating that the main results of this work are driven by the data rather than by the assumed mean function.

Also a key feature of GPs is that derivatives of a GP are also GPs.  
If $k$ is the kernel, then
\begin{align}
\mathrm{cov}\!\Big(f_i, \frac{\partial f_j}{\partial x_j}\Big)
  &= \frac{\partial k(x_i,x_j)}{\partial x_j}, \\
\mathrm{cov}\!\Big(\frac{\partial f_i}{\partial x_i}, \frac{\partial f_j}{\partial x_j}\Big)
  &= \frac{\partial^2 k(x_i,x_j)}{\partial x_i \partial x_j},
\end{align}
and the derivative prediction satisfies
\begin{equation}
\bar{\hat f'} 
= \hat\mu' + K'(X,\hat X)^{T} [K+C]^{-1} (y-\mu),
\end{equation}
with covariance
\begin{equation}
\mathrm{cov}(\hat f') 
= K''(\hat X,\hat X) - K'(X,\hat X)^{T} [K+C]^{-1} K'(X,\hat X).
\end{equation}

Thus both a function and its derivatives can be reconstructed consistently within the same statistical framework, allowing cosmological quantities depending on $H(z)$, $H'(z)$, and $H''(z)$ to be obtained by sampling from their joint GP distribution.
\subsection{Choice of covariance kernel}

The kernel encodes assumptions about the smoothness and differentiability of the reconstructed function. A widely used baseline is the Squared-Exponential (SE) kernel,
\begin{equation}
k_{\mathrm{SE}}(r)=\sigma^{2}\exp\!\left(-\frac{r^{2}}{2\ell^{2}}\right),
\label{eq:seqexpkernel}
\end{equation}
which is infinitely differentiable and often oversmooths cosmological reconstructions. For greater flexibility, we adopt the Matérn class,
\begin{equation}
k_{\nu}(r)=\sigma^{2}\frac{2^{1-\nu}}{\Gamma(\nu)}
\left(\frac{\sqrt{2\nu}\,r}{\ell}\right)^{\nu}
K_{\nu}\!\left(\frac{\sqrt{2\nu}\,r}{\ell}\right),
\label{eq:maternkernel}
\end{equation}
where $\sigma^{2}$ is the variance, $\ell$ the correlation length, $\nu$ controls smoothness, and $K_{\nu}$ is the modified Bessel function.  
For half-integer $\nu$, this reduces to closed forms useful for cosmology:
\begin{itemize}

\item \textbf{Matérn 3/2:}
\[
k_{3/2}(r)
= \sigma^{2}\!\left(1+\frac{\sqrt{3}r}{\ell}\right)
   e^{-\sqrt{3}r/\ell}.
\]

\item \textbf{Matérn 5/2:}
\[
k_{5/2}(r)
= \sigma^{2}\!\left(
1+\frac{\sqrt{5}r}{\ell}
 +\frac{5r^{2}}{3\ell^{2}}
\right)
e^{-\sqrt{5}r/\ell}.
\]

\item \textbf{Matérn 7/2:}
\[
k_{7/2}(r)
= \sigma^{2}\!\left(
1+\frac{\sqrt{7}r}{\ell}
 +\frac{14r^{2}}{5\ell^{2}}
 +\frac{7\sqrt{7}r^{3}}{15\ell^{3}}
\right)
e^{-\sqrt{7}r/\ell}.
\]

\item \textbf{Matérn 9/2:}
\[
k_{9/2}(r)
= \sigma^{2}\!\left(
1+\frac{3r}{\ell}
 +\frac{27r^{2}}{7\ell^{2}}
 +\frac{18r^{3}}{7\ell^{3}}
 +\frac{27r^{4}}{35\ell^{4}}
\right)
e^{-3r/\ell}.
\]

\end{itemize}

Kernel choice becomes crucial when reconstructing not only a function $f$ but also its derivatives \cite{Seikel:2012cs,Seikel:2013fda,Mukherjee:2022yyq}.  
In cosmology, many key quantities depend on $H(z)$, $H'(z)$, and $H''(z)$ \cite{Mukherjee:2024ryz,Favale:2023lnp}.  
Starting from the Hubble rate and the equation of state
\begin{equation}
H(z)=H_{0}\sqrt{\Omega_{m,0}(1+z)^{3}+(1-\Omega_{m,0})+\Omega_{k,0}(1+z)^{2}},
\end{equation}
\begin{equation}
w(z) = \frac{2(1+z)HH' - 3H^{2}}
            {3H^{2} - \Omega_{m,0}H_{0}^{2}(1+z)^{3}}.
\end{equation}
Beyond $H(z)$ and $w(z)$, the GP reconstruction naturally extends to other cosmological quantities such as the deceleration parameter $q(z)$ and the jerk $j(z)$ \cite{Mukherjee:2024ryz}, which depend on the first and second derivatives of $H(z)$. Since different kernels imply different degrees of derivative smoothness, their performance must be assessed directly on cosmological data. 
In this work, we reconstruct $H(z)$ using several Matérn kernels together with the Squared Exponential kernel, and examine their behaviour using cosmic chronometer data, with the robustness of the inferred trends further evaluated through a $\chi^2$ analysis incorporating recent DESI BAO DR2 measurements.

In the present work the Gaussian Process reconstruction of
$H(z)$ and its derivative $\left.\frac{dH}{dz}\right|_{z=0}$ is performed using the 32 cosmic chronometer (CC)
measurements only. The DESI DR2 baryon acoustic oscillation dataset is incorporated separately as an auxiliary dataset for statistical comparison of the kernel-dependent $\chi^2$
behaviour. In particular, we use the six  BAO
measurements of $D_H/r_d$ reported by the DESI Collaboration\cite{DESI:2025zgx}.
These are converted to effective Hubble parameter constraints
through
\begin{equation}
 H(z)=\frac{c}{(D_H/r_d)\,r_d}   
\end{equation}
adopting a fixed sound horizon $r_d = 147\,{\rm Mpc}$ \cite{Zhang:2025bmk}.
The corresponding DESI covariance matrix is included when
evaluating the combined $\chi^2$ statistics.
\subsection{Tree approach}
\begin{figure*}[!t]
\centering

\begin{minipage}{0.49\textwidth}
    \centering
    \includegraphics[width=\linewidth]{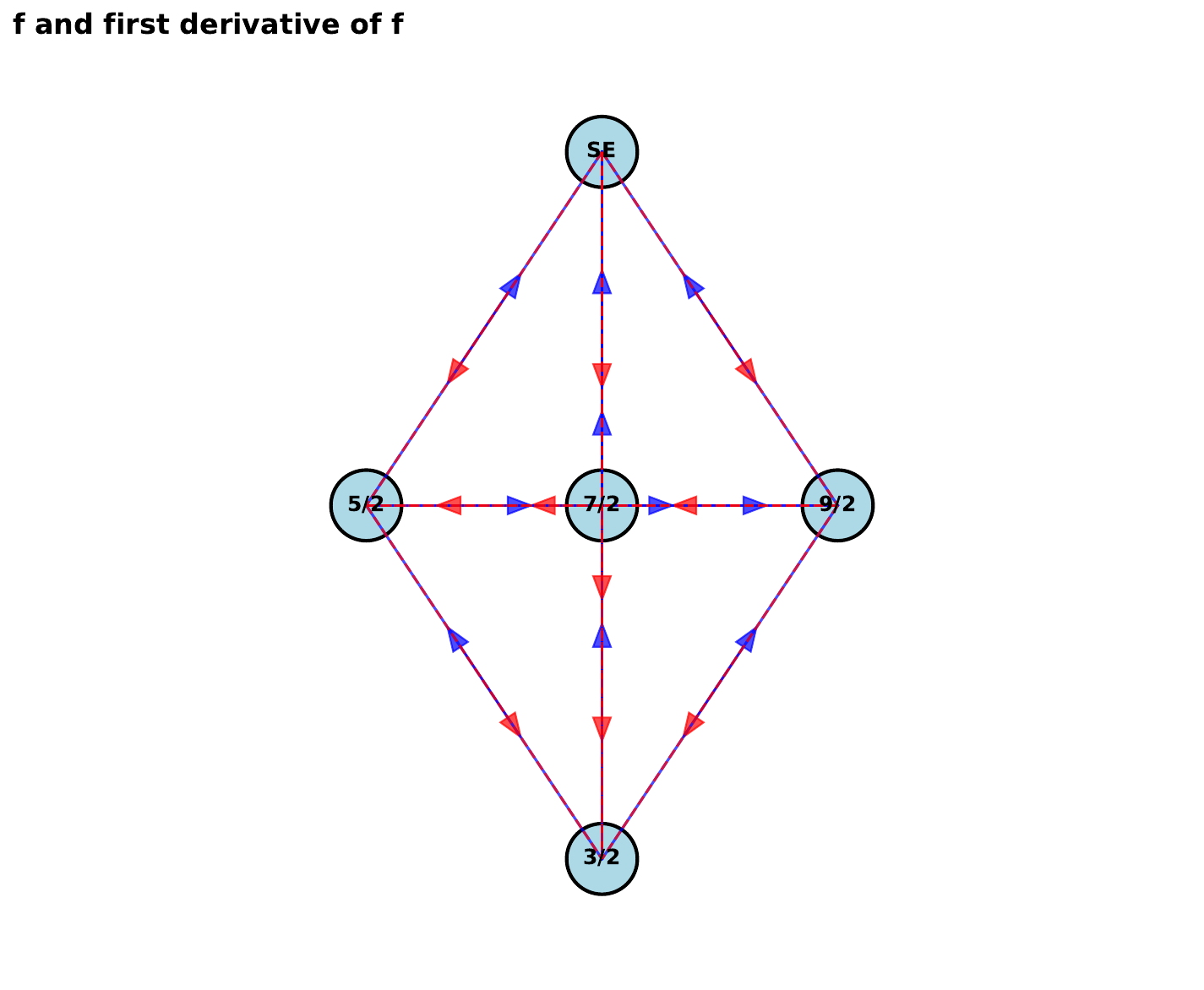}
    \vspace{2pt}
    {\small (a)}
\end{minipage}
\hfill
\begin{minipage}{0.49\textwidth}
    \centering
    \includegraphics[width=\linewidth]{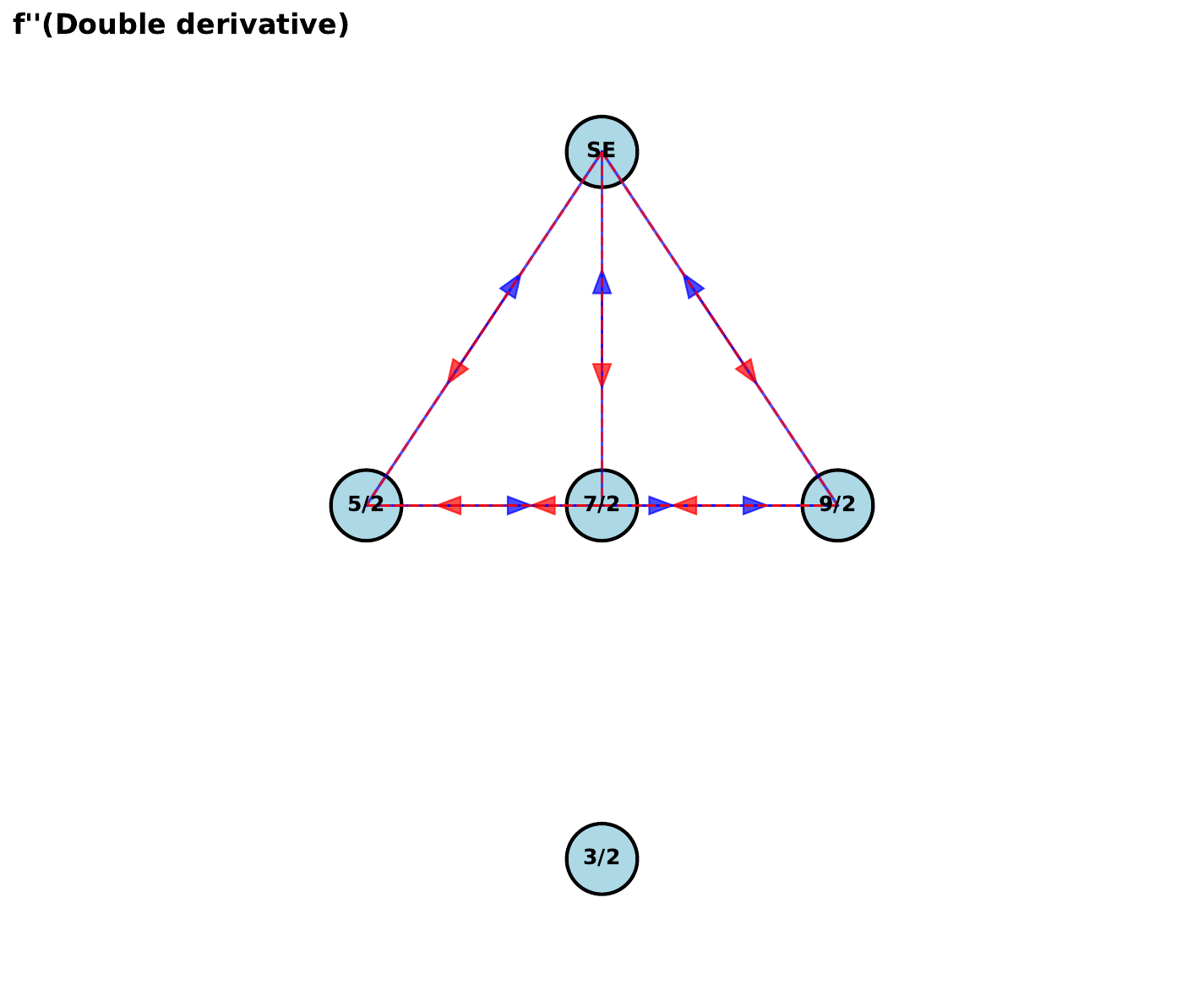}
    \vspace{2pt}
    {\small (b)}
\end{minipage}

\vspace{8pt}

\begin{minipage}{0.49\textwidth}
    \centering
    \includegraphics[width=\linewidth]{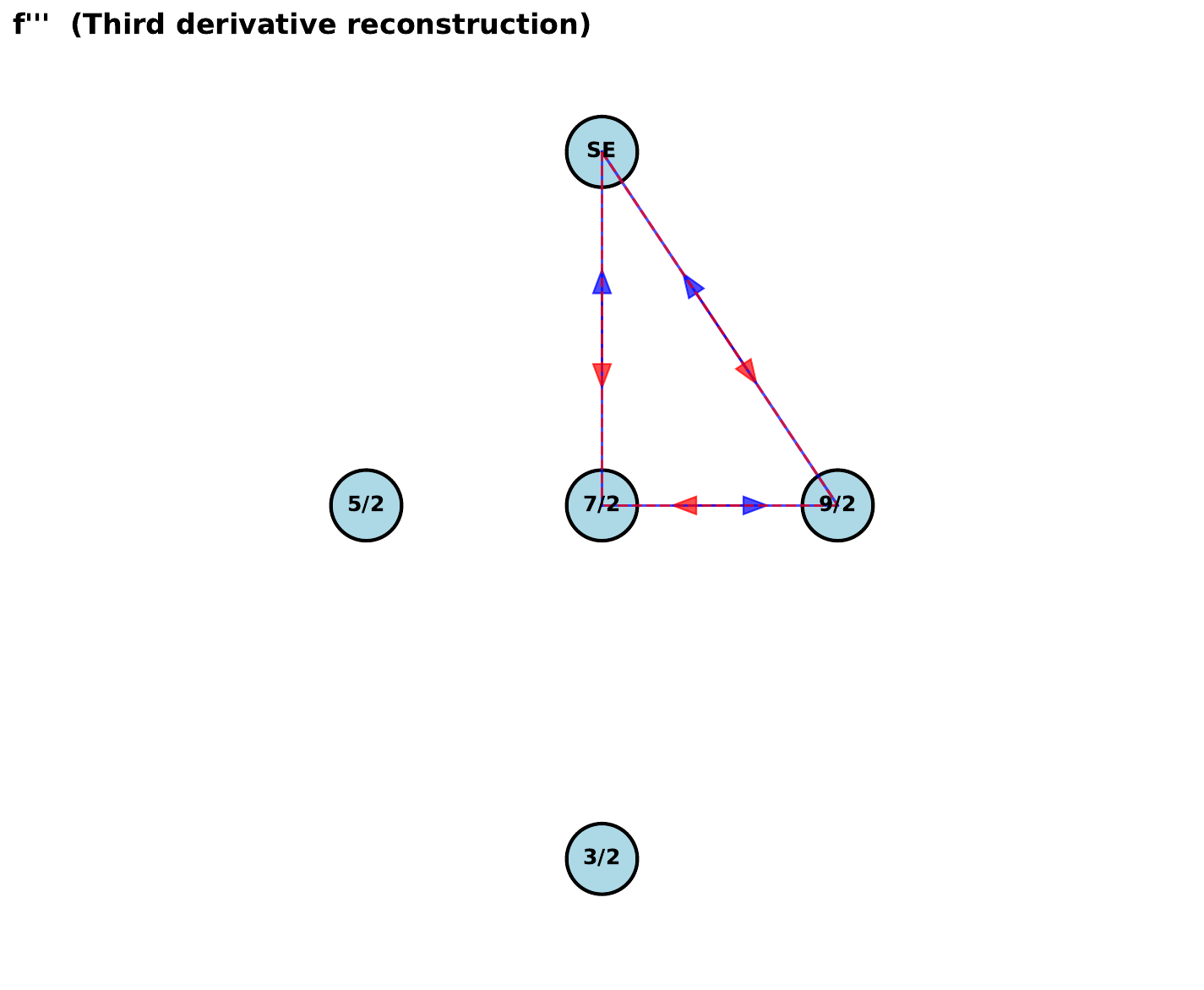}
    \vspace{2pt}
    {\small (c)}
\end{minipage}
\hfill
\begin{minipage}{0.49\textwidth}
    \centering
    \includegraphics[width=\linewidth]{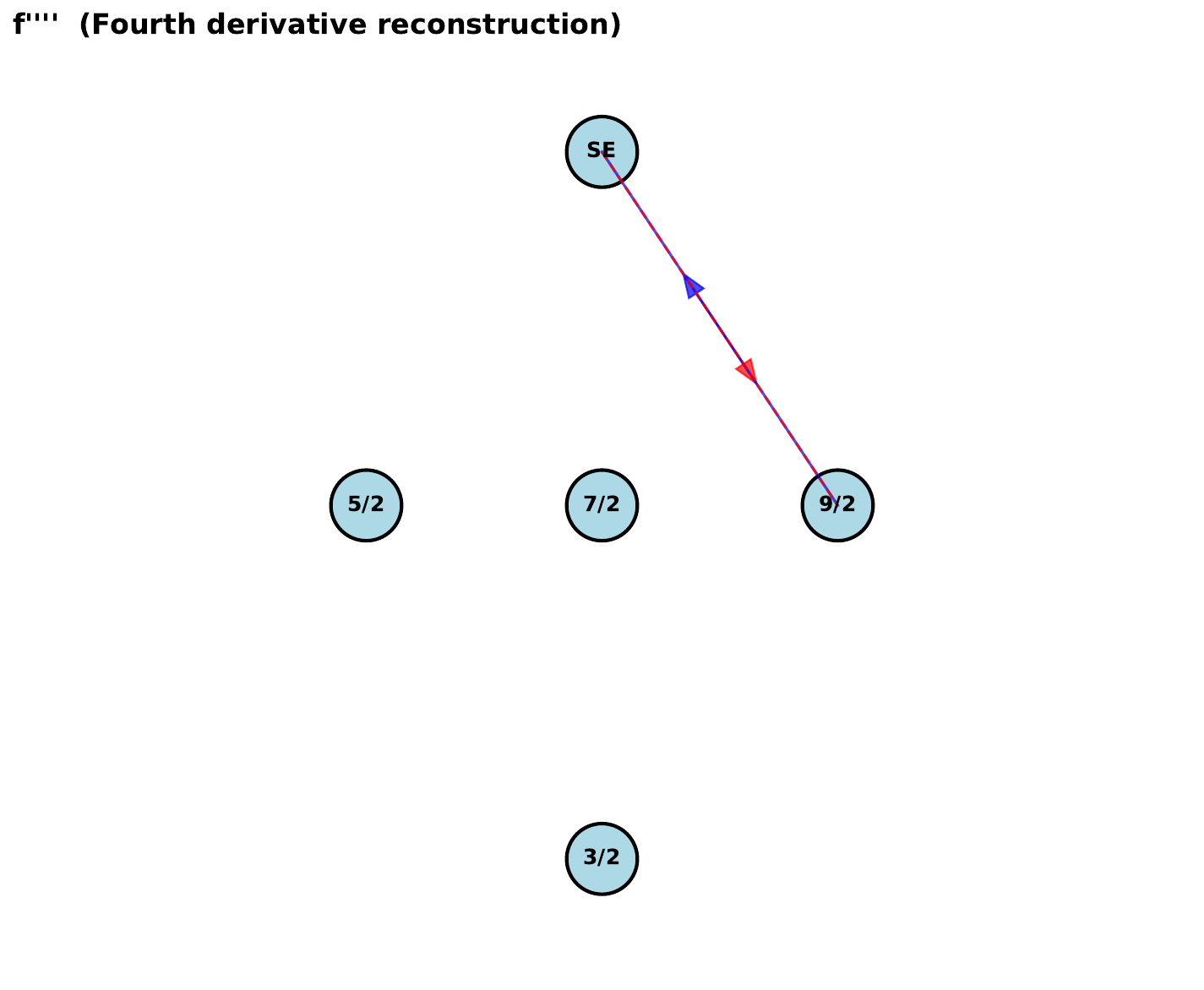}
    \vspace{2pt}
    {\small (d)}
\end{minipage}

\caption{
Illustration of the kernel-dependent derivative hierarchy in Gaussian Process reconstruction. 
Panel (a) shows the reconstruction of the function $f$ and its first derivative $f'$, which can be consistently obtained using all kernels considered (Matérn $3/2$, $5/2$, $7/2$, $9/2$, and Squared Exponential). 
Panel (b) displays the reconstruction of the second derivative $f''$, for which the Matérn $3/2$ kernel is excluded due to insufficient differentiability. 
Panel (c) shows the third derivative $f'''$, where both Matérn $3/2$ and $5/2$ kernels are eliminated. 
Panel (d) presents the fourth derivative $f''''$, which can only be reconstructed using the Matérn $9/2$ and Squared Exponential kernels. 
This figure visualizes the progressive elimination of kernels as higher-order derivatives are required, highlighting the role of kernel smoothness in cosmological reconstructions.
}\label{fig:treeplot}
\end{figure*}

As discussed earlier, we focus on the Squared-Exponential kernel and the Matérn family, 
specifically the Matérn $3/2$, $5/2$, $7/2$, and $9/2$ forms. 
The choice of kernel is crucial, as each kernel encodes a distinct level of smoothness 
and therefore imposes different priors on the reconstructed function and its derivatives. 
To systematize these choices, we introduce a tree-based scheme (Fig.~\ref{fig:treeplot}), 
which organizes the admissible kernels according to the highest derivative that can be 
consistently reconstructed.

The tree approach is particularly relevant when one aims to \emph{compare results across different kernels}. 
In such cases, moving hierarchically from Matérn $3/2$ to progressively smoother kernels 
(Matérn $5/2$, $7/2$, $9/2$, and ultimately the Squared-Exponential) allows for a controlled 
assessment of how increasing smoothness affects both the reconstructed quantities and the 
associated hyperparameters $(\ell,\sigma_f)$(Fig.~\ref{fig:trend}). 
The blue and red arrows shown in (Fig.~\ref{fig:treeplot}) illustrate possible transitions between kernels; 
however, these paths are not unique or mandatory and are included solely to guide systematic comparisons. 
If one performs a reconstruction using a single kernel only, the tree approach does not impose 
any additional constraint and has no direct impact on the results.

The key organizing principle of the tree approach is the differentiability of the kernel. 
As shown in the top-left panel Fig.~\ref{fig:treeplot}, all kernels permit the reconstruction of the function $f$ 
and its first derivative $f'$. 
However, the absence of outgoing arrows from the Matérn $3/2$ kernel indicates that 
second derivatives cannot be consistently obtained with this kernel. 
At the next level, both the Matérn $3/2$ and $5/2$ kernels are excluded when reconstructing 
third derivatives, while only the Matérn $9/2$ and Squared-Exponential kernels remain viable 
for fourth-order derivatives. 
For derivatives of order higher than four, the Squared-Exponential kernel is the only admissible choice 
due to its infinite differentiability. 
This hierarchy ensures that kernel selection is dictated primarily by the derivative order required 
by the physical problem, rather than by arbitrary preference.
\section{Hubble reconstruction by different kernels
}
\begin{figure*}[t]
\centering

\begin{minipage}{0.49\textwidth}
    \centering
    \includegraphics[width=\linewidth]{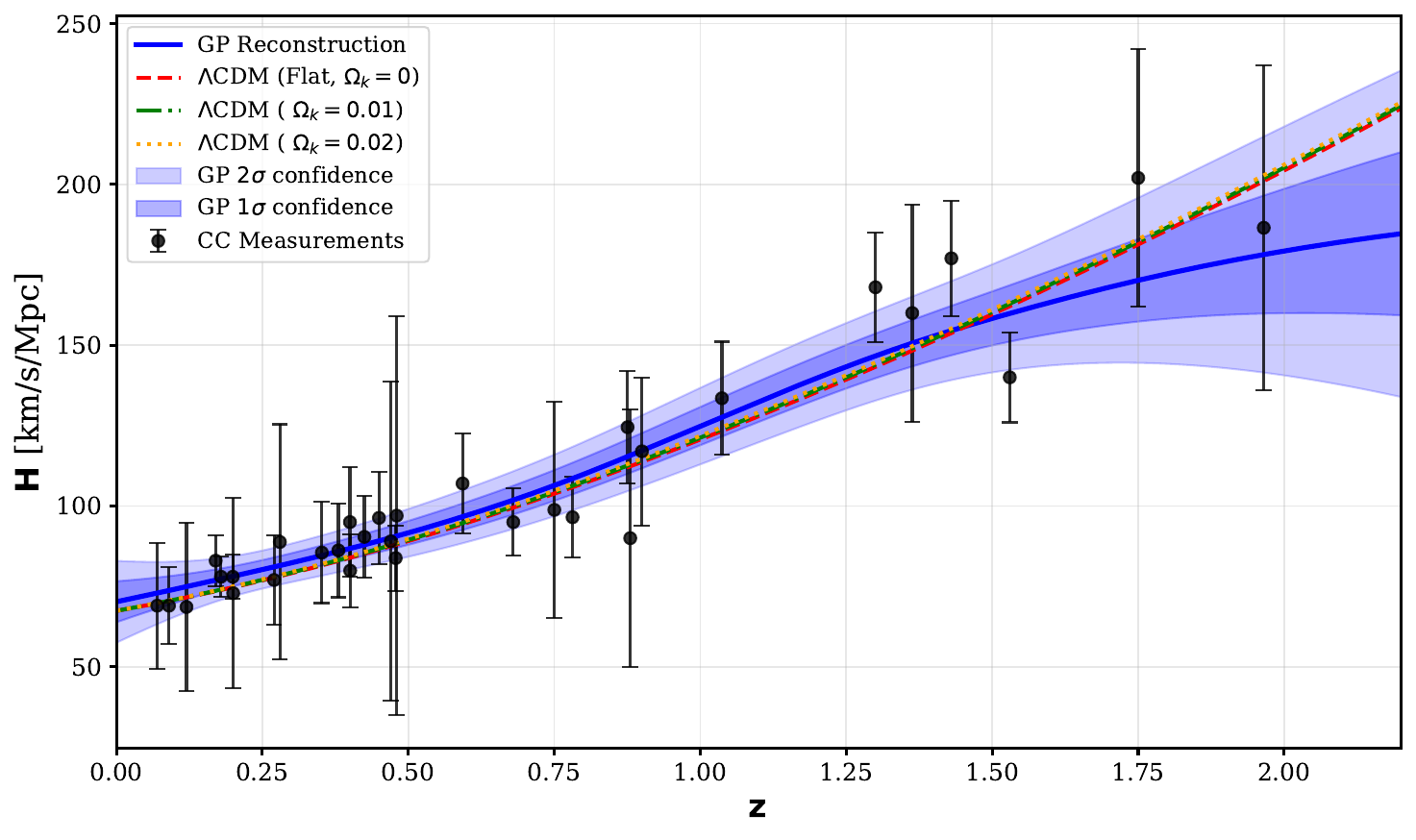}
    \vspace{2pt}
    {\small (a)}
\end{minipage}
\hfill
\begin{minipage}{0.49\textwidth}
    \centering
    \includegraphics[width=\linewidth]{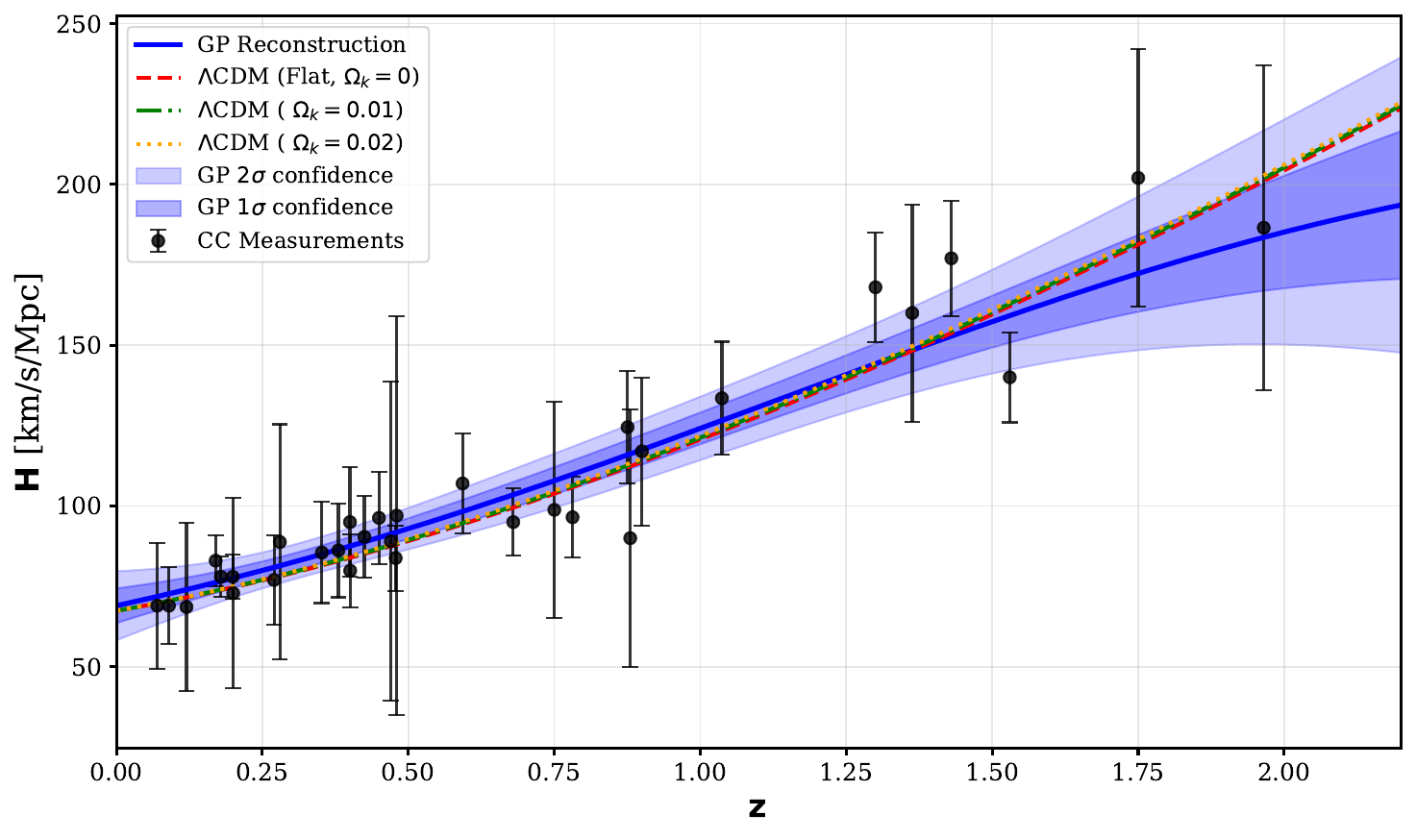}
    \vspace{2pt}
    {\small (b)}
\end{minipage}

\vspace{8pt}

\begin{minipage}{0.49\textwidth}
    \centering
    \includegraphics[width=\linewidth]{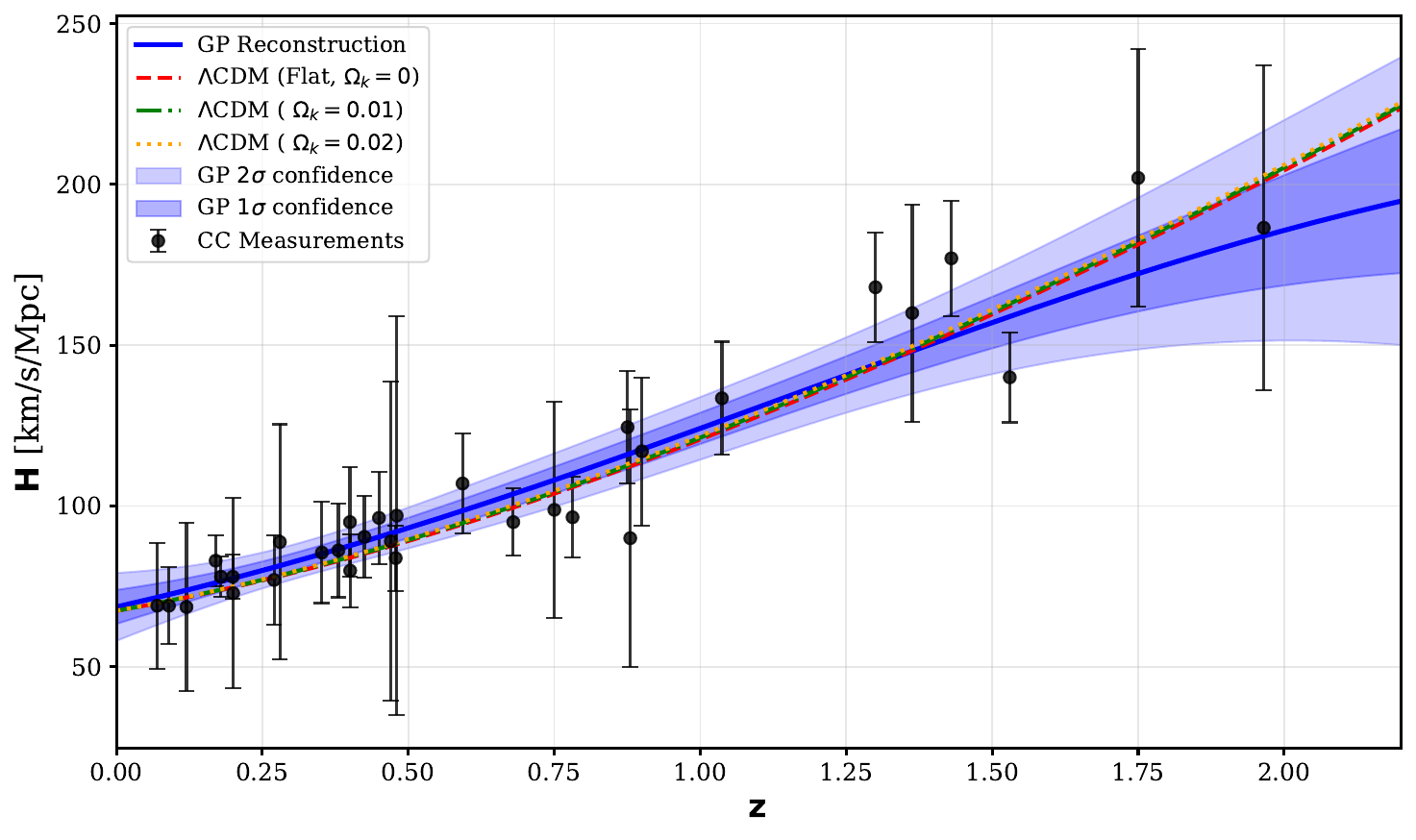}
    \vspace{2pt}
    {\small (c)}
\end{minipage}
\hfill
\begin{minipage}{0.49\textwidth}
    \centering
    \includegraphics[width=\linewidth]{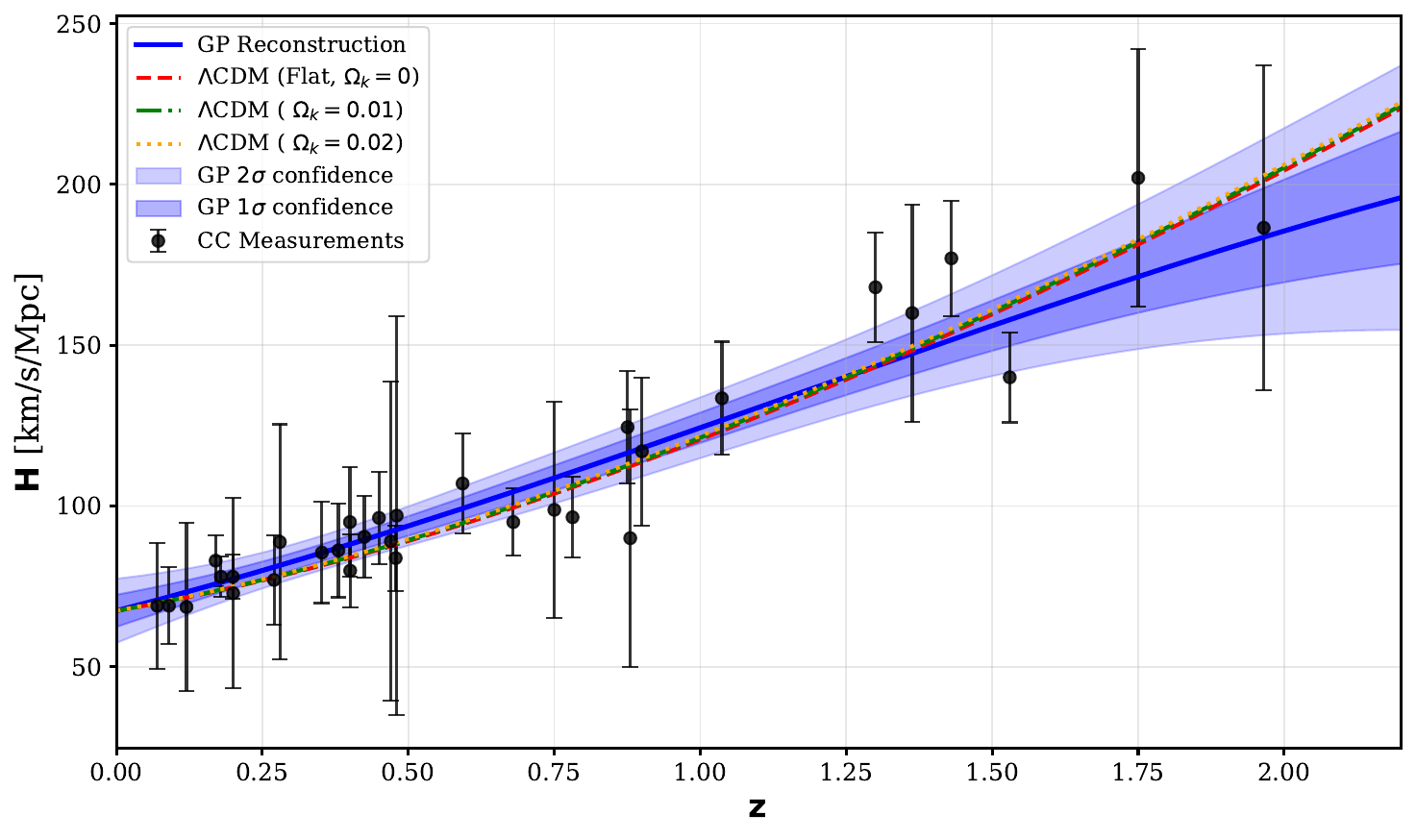}
    \vspace{2pt}
    {\small (d)}
\end{minipage}

\caption{
Gaussian Process reconstruction of the Hubble parameter $H(z)$ from 32 cosmic chronometer measurements using different covariance kernels. 
Panels (a)–(d) correspond to the Matérn $3/2$, Matérn $7/2$, Matérn $9/2$, and Squared Exponential kernels, respectively. 
In each panel, the solid blue curve denotes the GP mean reconstruction, with the shaded bands indicating the $1\sigma$ and $2\sigma$ confidence regions, while black points represent the CC data. 
The GP reconstructions are compared with $\Lambda$CDM models for different spatial curvature values ($\Omega_k=0$, $0.01$, and $0.02$).}
\label{fig:Hrec}
\end{figure*}

\begin{figure}[tp]
\centering
\includegraphics[width=1.0\linewidth]{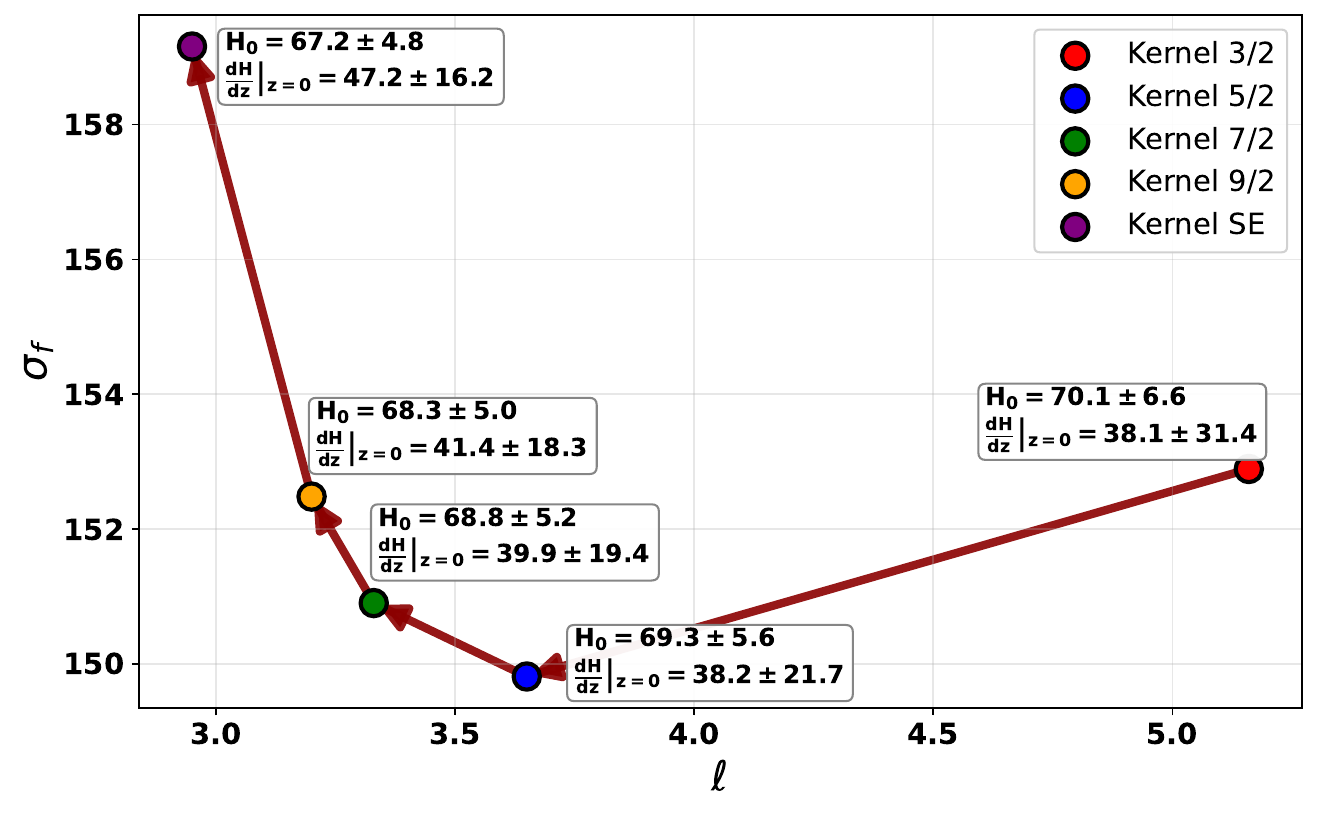}
\caption{
Hyperparameter space $(\ell,\sigma_f)$ of Gaussian Process kernels. 
Colored points correspond to the Matérn-$3/2$, $5/2$, $7/2$, $9/2$, and squared--exponential (SE) kernels, with arrows indicating increasing kernel smoothness. 
Each point is annotated with the reconstructed $H_0$ and the derivative $\left.\frac{dH}{dz}\right|_{z=0}$ (both with $1\sigma$ uncertainties), illustrating their dependence on the kernel choice.
}
\label{fig:trend}
\end{figure}

In this section, we reconstruct the Hubble parameter $H(z)$ from 32 cosmic-chronometer measurements using five Gaussian-Process kernels (Matérn $3/2$, $5/2$, $7/2$, $9/2$, and Squared Exponential) and compare them with $\Lambda$CDM models of different curvature Fig.~\ref{fig:Hrec}. A clear trend emerges: as the kernel smoothness increases from Matérn $3/2$ to the infinitely differentiable Squared Exponential, the inferred $H_{0}$ decreases monotonically from $70.1 \pm 6.6$ km\,s$^{-1}$Mpc$^{-1}$ to $67.2 \pm 4.8$ km\,s$^{-1}$Mpc$^{-1}$, with uncertainties shrinking accordingly. This nearly 3 km\,s$^{-1}$Mpc$^{-1}$ shift highlights the sensitivity of GP reconstructions to the smoothness prior encoded in the covariance kernel. The numerical values of the optimized hyperparameters and the reconstructed 
quantities are summarized in Table~\ref{tab:kernel_results}. For clarity, Fig.~\ref{fig:trend} visualizes these 
results in the hyperparameter space $(\ell,\sigma_f)$, illustrating how the 
reconstructed $H_0$ and the derivative $\left.\frac{dH}{dz}\right|_{z=0}$ vary 
systematically across kernels with increasing smoothness. The arrows indicate 
the direction of increasing kernel smoothness from the Matérn-$3/2$ kernel to 
the squared--exponential kernel.

The Matérn sequence provides a smoothness hierarchy that directly controls the rigidity of the reconstructed $H(z)$. The Matérn–$3/2$ kernel, being the least smooth, preserves local variations in the CC data and yields a steeper slope near $z\simeq 0$, producing higher $H_{0}$ values consistent with SH0ES like\cite{Riess:2021jrx}. Increasing $\nu$ to $5/2$, $7/2$, and $9/2$ progressively suppresses small-scale features, flattens $H(z)$ near the origin, and shifts $H_{0}$ downward. In the Squared Exponential limit, the reconstruction becomes extremely smooth and nearly insensitive to local CC fluctuations, leading to the lowest $H_{0}$ values, close to Planck like \cite{Planck:2018vyg}. Thus, the kernel hierarchy yields a continuous ordering—
 \emph{Matérn--$3/2$ $\rightarrow$ highest $H_{0}$}, \emph{Matérn--$5/2$ $\rightarrow$ slightly lower}, \emph{Matérn--$7/2$ and $9/2$ $\rightarrow$ progressively lower}, and \emph{Squared Exponential $\rightarrow$ lowest},
spanning the range between the SH0ES and Planck determinations of $H_{0}$ in a fully non-parametric manner.
\begin{table*}[t]
\centering
\caption{Gaussian Process hyperparameters, reconstructed values of the Hubble constant $H_0$, its first derivative at $z=0$, and log marginal likelihood (LML) for different covariance kernels with uncertainties correspond to $1\sigma$ GP errors. We have used CC 32 and 6 $DH/r_{d}$ DESI DR2 measurements.}
\setlength{\tabcolsep}{5pt}
\renewcommand{\arraystretch}{1.2}
\begin{tabular}{lcccccc}
\hline\hline
\textbf{Kernel} 
& $\boldsymbol{\sigma_f}$ 
& $\boldsymbol{l}$ 
& $\boldsymbol{H_0}$ (km\,s$^{-1}$\,Mpc$^{-1}$) 
& $\boldsymbol{\left.\dfrac{dH}{dz}\right|_{z=0}}$ (km\,s$^{-1}$\,Mpc$^{-1}$)
& $\boldsymbol{\mathrm{LML\ (CC)}}$ 
& $\boldsymbol{\mathrm{LML\ (CC+DESI\ DR2)}}$ \\
\hline

Matérn-$3/2$        & 152.89 & 5.16 & $70.1 \pm 6.6$ & $38.1 \pm 31.4$ & $-137.01$ & $-155.29  $ \\
Matérn-$5/2$        & 149.81 & 3.65 & $69.3 \pm 5.6$ & $38.2 \pm 21.7$ & $-136.96$ & $-154.35$ \\
Matérn-$7/2$        & 150.90 & 3.33 & $68.8 \pm 5.2$ & $39.9 \pm 19.4$ & $-136.99$ & $-154.21 $ \\
Matérn-$9/2$        & 152.48 & 3.20 & $68.3 \pm 5.0$ & $41.4 \pm 18.3$ & $-137.02$ & $-154.18 $ \\
Squared Exponential & 159.16 & 2.95 & $67.2 \pm 4.8$ & $47.2 \pm 16.2$ & $-137.09 $ & $-154.19 $ \\

\hline\hline
\end{tabular}
\label{tab:kernel_results}
\end{table*}
This behavior reflects the balance GP methods impose between data fidelity and smoothness priors means flexible kernels track the CC data closely, while smoother kernels underfit local structure and drive $H_{0}$ toward lower values.
\subsection{Kernel Differentiability and Derivative Reconstruction}

In addition to the zeroth-order quantity $H_0$, Gaussian Process regression allows a direct reconstruction of derivative-level observables within the same statistical framework. Since the central claim of this work concerns the role of kernel differentiability, it is necessary to verify that the hierarchy manifests explicitly in derivative quantities and not solely in $H(z)$.

\begin{figure*}[t]
\centering

\begin{minipage}{0.49\textwidth}
    \centering
    \includegraphics[width=\linewidth]{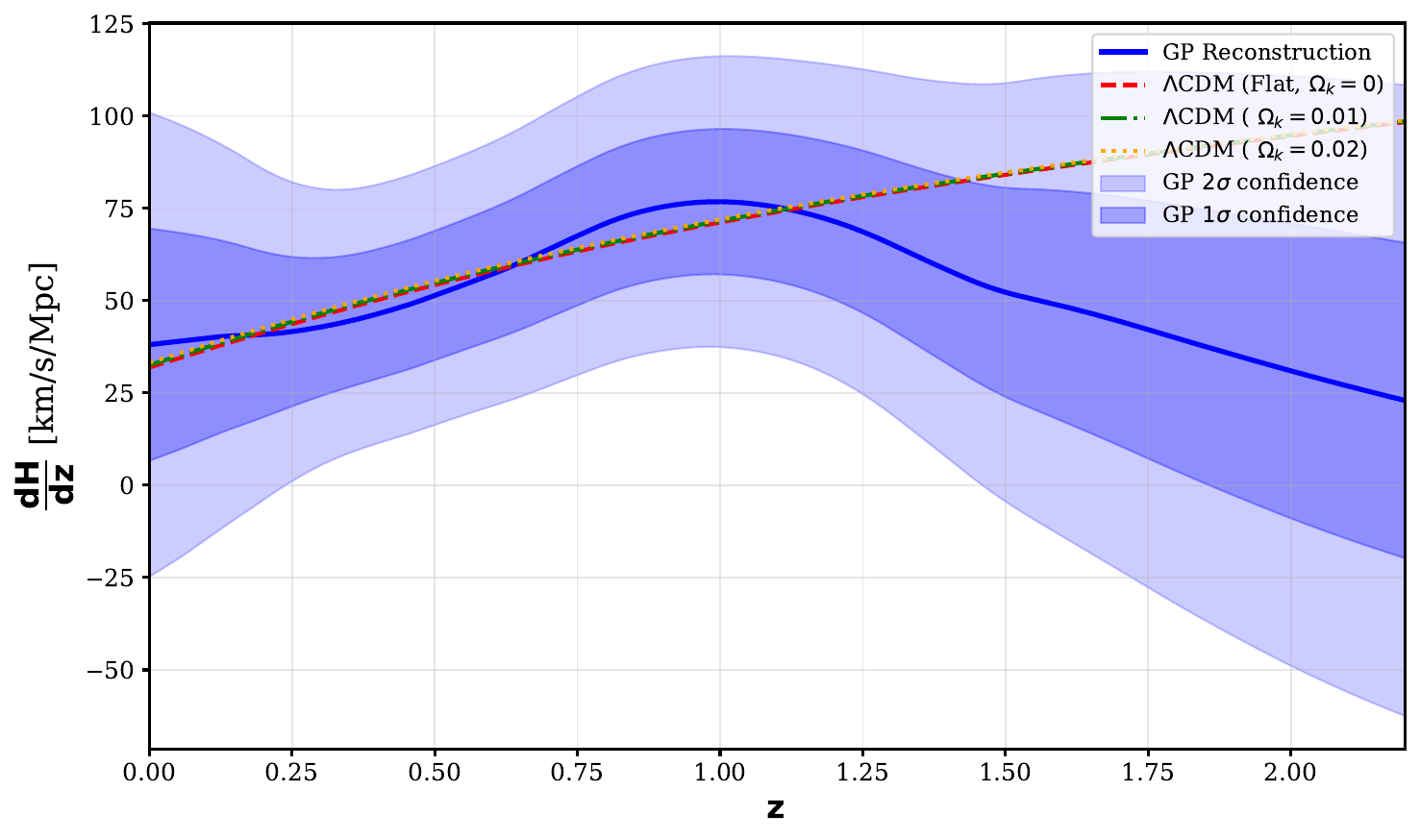}
    \vspace{2pt}
    {\small (a)}
\end{minipage}
\hfill
\begin{minipage}{0.49\textwidth}
    \centering
    \includegraphics[width=\linewidth]{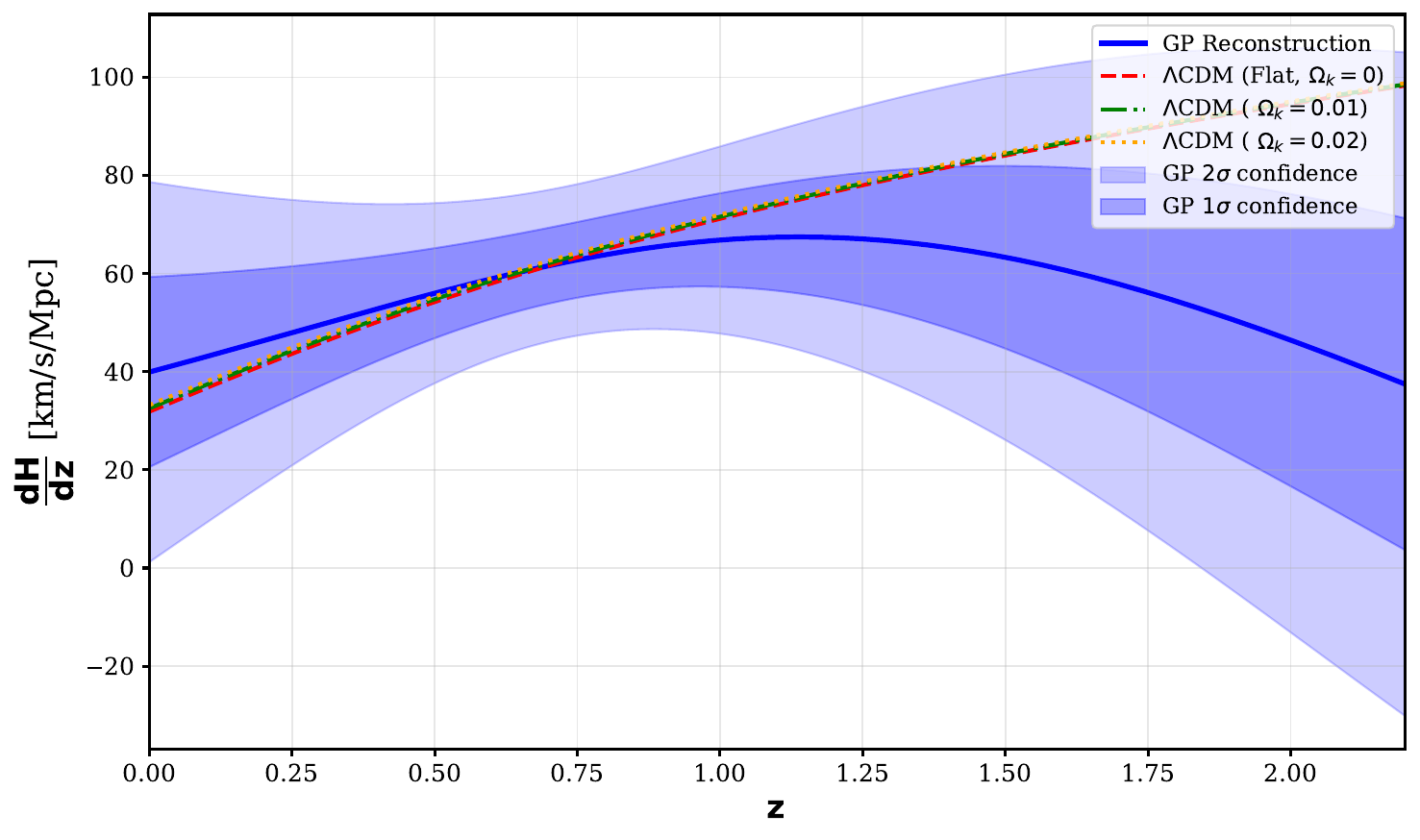}
    \vspace{2pt}
    {\small (b)}
\end{minipage}

\vspace{8pt}

\begin{minipage}{0.49\textwidth}
    \centering
    \includegraphics[width=\linewidth]{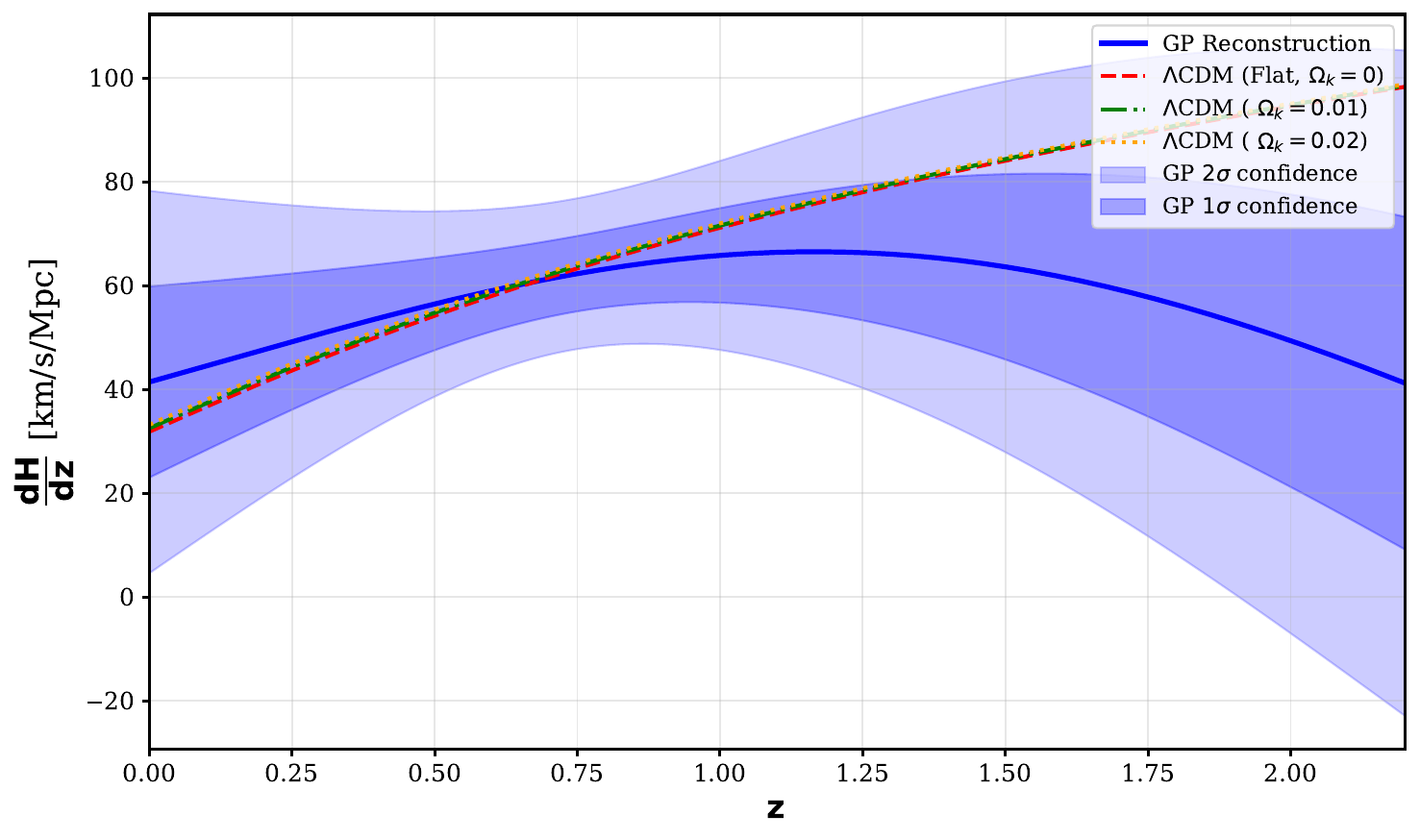}
    \vspace{2pt}
    {\small (c)}
\end{minipage}
\hfill
\begin{minipage}{0.49\textwidth}
    \centering
    \includegraphics[width=\linewidth]{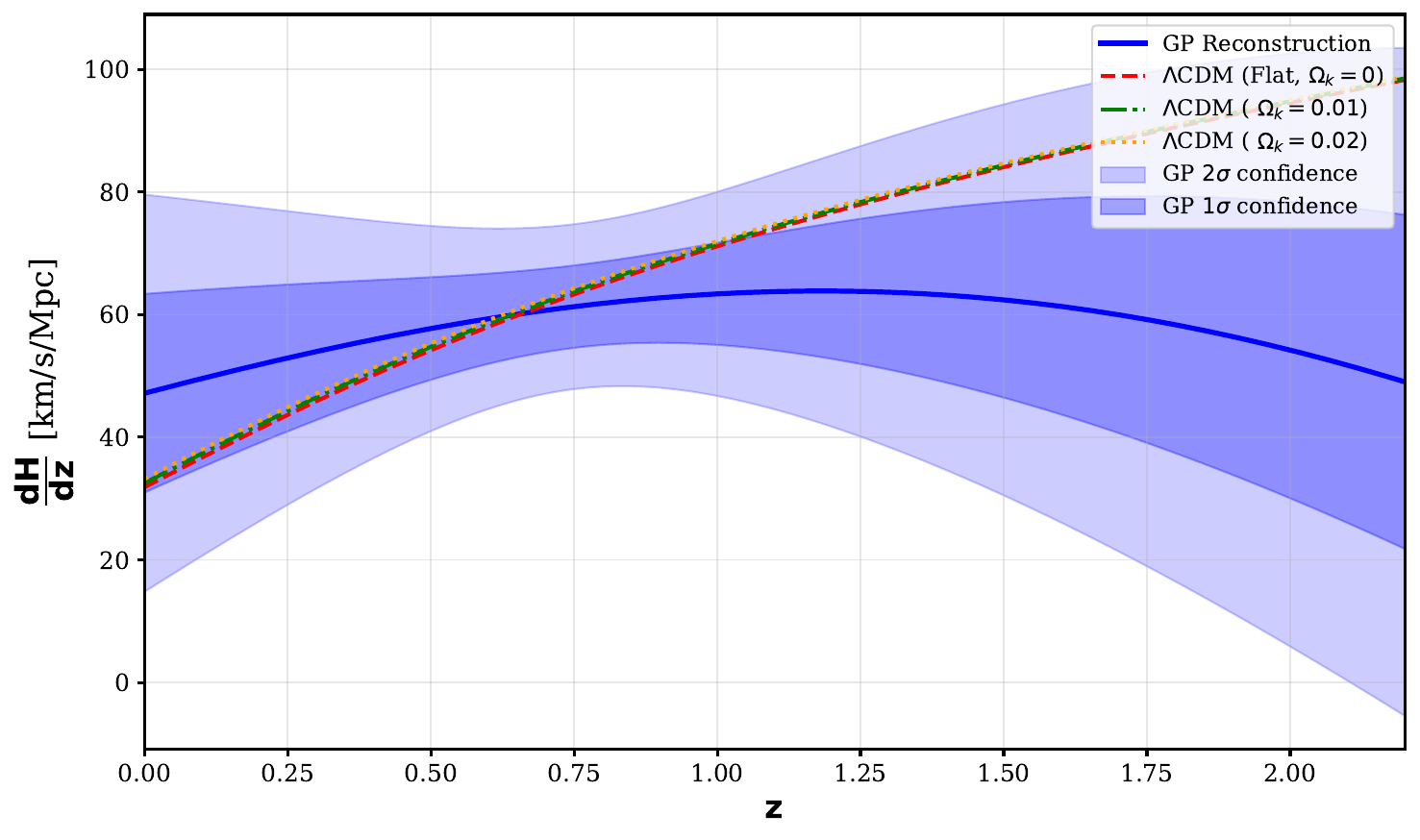}
    \vspace{2pt}
    {\small (d)}
\end{minipage}

\caption{
Gaussian Process reconstruction of the first derivative of the Hubble parameter, $dH/dz$, from 32 cosmic chronometer measurements using different covariance kernels. 
Panels (a)–(d) correspond to the Matérn $3/2$, Matérn $7/2$, Matérn $9/2$, and Squared Exponential kernels, respectively. 
In each panel, the solid blue curve denotes the GP mean reconstruction, with the shaded bands indicating the $1\sigma$ and $2\sigma$ confidence regions. 
The derivative reconstructions are compared with the corresponding $\Lambda$CDM predictions for different spatial curvature values ($\Omega_k=0$, $0.01$, and $0.02$).
}
\label{fig:Hprime}
\end{figure*}

Table~\ref{tab:kernel_results} reports the reconstructed first derivative at the present epoch,
$\left.\frac{dH}{dz}\right|_{z=0}$, for each kernel. The same trend is also
visualized in Fig.~\ref{fig:trend}, where the hyperparameter space $(\ell,\sigma_f)$
illustrates how the reconstructed $H_0$ and $\left.\frac{dH}{dz}\right|_{z=0}$
vary systematically along the kernel smoothness hierarchy.
A clear monotonic trend is observed across the smoothness hierarchy. 
As the differentiability increases from Matérn-$3/2$ to the Squared Exponential kernel, 
the reconstructed slope at $z=0$ increases systematically. Thus, progressively smoother kernels enforce a steeper local slope in the expansion history at low redshift Fig.~\ref{fig:Hprime}.

Importantly, this monotonic increase in $H'(0)$ coincides with a monotonic decrease in the inferred $H_0$. Rougher kernels, which allow greater local flexibility, yield smaller slopes at $z=0$ and correspondingly higher extrapolated values of $H_0$, whereas smoother kernels impose stronger derivative regularity, increasing $H'(0)$ while shifting $H_0$ toward lower values. The quantitative anti-correlation between $H'(0)$ and $H_0$ is consistent with the interpretation that differentiability constraints propagate into zeroth-order inference. Fig.~\ref{fig:trend}.

This behaviour follows from the structure of the covariance kernels. The Matérn family restricts the order of admissible derivatives through its smoothness parameter $\nu$, while the Squared Exponential kernel enforces infinite differentiability. As smoothness increases, small-scale variations in $H(z)$ are progressively suppressed, modifying the local slope near the origin and thereby producing the observed hierarchy in $H_0$.

The derivative-level analysis therefore shows that the kernel hierarchy is not merely interpretive, but manifests quantitatively in the reconstructed first derivative and provides a consistent explanation for the ordered shifts in $H_0$.
\subsection{$\chi^2$ and LML Analysis}
The behaviour of the goodness of fit, quantified by the chi-square statistic 
$\chi^{2}$, also exhibits a clear and theoretically well-motivated dependence 
on the smoothness of the Gaussian Process kernel \cite{Velazquez:2024aya}. For a dataset 
$\{z_i, H_i, \sigma_i\}$, the GP prediction at the data locations is 
\begin{equation}
\boldsymbol{\mu} = K \left( K + N \right)^{-1} \mathbf{H},
\end{equation}
where $K$ is the kernel matrix with elements 
$K_{ij} = k(|z_i - z_j|)$ and $N$ denotes the total observational covariance matrix of the cosmic chronometer data, incorporating both statistical and systematic contributions. The systematic component accounts for effects such as stellar population
synthesis (SPS) modeling and stellar libraries, which may
introduce correlations between measurements at different
redshifts. The analysis therefore employs the full non-diagonal covariance matrix provided by Moresco et al.~(2020)\cite{Moresco:2020fbm} in order to
account for these correlated systematic contributions. The effect of correlated systematics on the reconstructed
uncertainties is modest, while the monotonic kernel hierarchy in $H_0$ reported in Table~\ref{tab:kernel_results}
remains clearly visible.
\begin{equation}
 N = C^{\rm stat} + C^{\rm sys} 
 \label{eq:16}
\end{equation}

The residual vector is $\mathbf{d} = \mathbf{H} - \boldsymbol{\mu}$,
and the chi-square statistic is written as
\begin{equation}
\chi^2 = \mathbf{d}^T N^{-1} \mathbf{d}.
\label{eq:chisq}
\end{equation}

The key point is that different kernels produce different kernel matrices $K$, 
which in turn modify the posterior mean $\boldsymbol{\mu}$ and therefore the 
residuals. The Squared Exponential (SE) kernel Equation~\eqref{eq:seqexpkernel} decays as $\exp(-r^{2})$, which is much faster, whereas the Matérn kernels  Equation~\eqref{eq:maternkernel}
decay only as $\exp(-r)$ inside the Bessel function. As a result, for typical CC 
separations $\Delta z \sim 0.05$--$0.3$, the SE kernel suppresses off-diagonal 
covariances $K_{ij}$ far more strongly than the Matérn-$3/2$ kernel:
\begin{equation}
K_{ij}^{(3/2)} \gg K_{ij}^{(\rm SE)} 
\quad \text{for the same } r = |z_i - z_j|.
\end{equation}

This difference directly affects the posterior mean. A rapidly decaying kernel 
(like SE) enforces a very smooth function and effectively averages over 
neighbouring points, making the reconstruction less responsive to local 
fluctuations in the CC data.
Consequently, the posterior mean $\mu_i$ lies 
farther from the observed $H_i$ values, producing large residuals 
$|d_i| = |H_i - \mu_i|$ and hence
\begin{equation}
\chi^2_{\rm SE} = \mathbf{d}_{\rm SE}^T N^{-1} \mathbf{d}_{\rm SE} 
 \quad \text{ is large.}
\end{equation}

In contrast, the Matérn-$3/2$ kernel decays more slowly, retains stronger 
correlations between mildly separated points, and allows sufficient flexibility 
for the reconstructed $H(z)$ to follow the local structure of the CC data. 
This produces smaller residuals as $|d_i|_{\,(3/2)} < |d_i|_{\rm (SE)},$ and therefore
\begin{equation}
\chi^2_{\,(3/2)} 
= \mathbf{d}_{(3/2)}^T N^{-1} \mathbf{d}_{(3/2)}
\quad \text{is significantly smaller}.
\end{equation}

However, it is important to emphasize that the $\chi^2$ ordering alone does not constitute a decisive model-selection criterion, since the GP hyperparameters are optimized on the same dataset. A training-residual $\chi^2$ primarily reflects differences in effective flexibility rather than predictive preference. To assess whether the data statistically favor any kernel, we compare the maximized log marginal likelihood (LML) obtained during hyperparameter optimization (Table~\ref{tab:kernel_results}). For the CC-only dataset the LML spread is extremely small ($\Delta \mathrm{LML} < 0.12$), and it remains modest even after including DESI DR2 measurements ($\Delta \mathrm{LML} \lesssim 1$). Here $\Delta \mathrm{LML}$ denotes the difference relative to the maximum LML across the tested kernels. These differences do not constitute decisive Bayesian evidence for one kernel over another. The observed $\chi^2$ hierarchy should therefore be interpreted descriptively: it quantifies how smoothness priors reshape the reconstruction rather than establishing a uniquely preferred kernel. The stability of the LML ranking under dataset expansion demonstrates that the main conclusions are robust against reasonable kernel choices.

\begin{figure}[tp]
\centering
\includegraphics[width=1.0\linewidth]{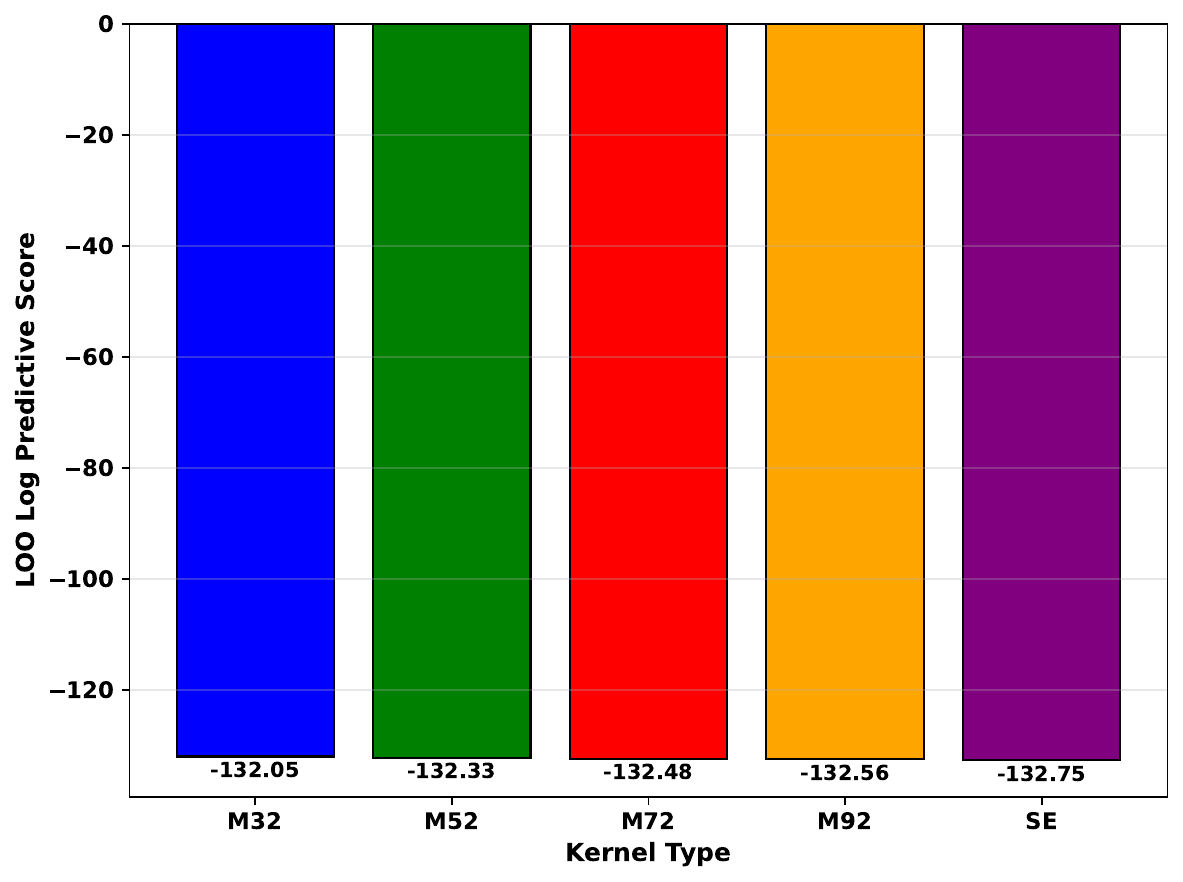}
\caption{Leave-one-out cross-validation (LOO-CV) predictive comparison for Gaussian Process kernels using the CC-32 dataset. The small variation in LOO scores ($\Delta \mathrm{LOO} \ll 1$) is evident.}
\label{fig:loo}
\end{figure}

To complement the LML analysis, we perform a leave-one-out cross-validation (LOO-CV) test as an additional check of predictive performance using the CC-32 dataset. The resulting LOO-CV scores show only minor differences across the kernel hierarchy, with $\Delta \mathrm{LOO} \ll 1$, indicating that no kernel is statistically preferred. This is consistent with the LML results and supports the interpretation that the observed kernel-dependent ordering reflects prior sensitivity rather than in-sample fitting effects. The comparison is shown in Fig.~\ref{fig:loo}.

The GP hyperparameters are obtained by maximizing the log
marginal likelihood using the truncated Newton (TNC)
optimizer implemented in the \texttt{GaPP} package.
The covariance amplitude and correlation length are
parameterized as strictly positive quantities and optimized within broad numerical bounds to avoid unphysical regions of parameter space. To verify robustness against local maxima, the optimization
was repeated using multiple independent initializations,
including $\theta = [50,1]$, $[100,0.5]$, and $[200,2]$,
effectively performing several random restarts of the
likelihood maximization. All runs converge to the same
maximum within numerical tolerance. The resulting
hyperparameters and kernel ordering are therefore stable
with respect to initialization and optimizer settings,
confirming that the LML comparison reflects genuine
statistical structure rather than optimization artifacts. This confirms that the kernel hierarchy discussed in this work reflects structured prior sensitivity rather than statistical model selection.

\begin{figure}[t]
\centering

\includegraphics[width=0.9\linewidth]{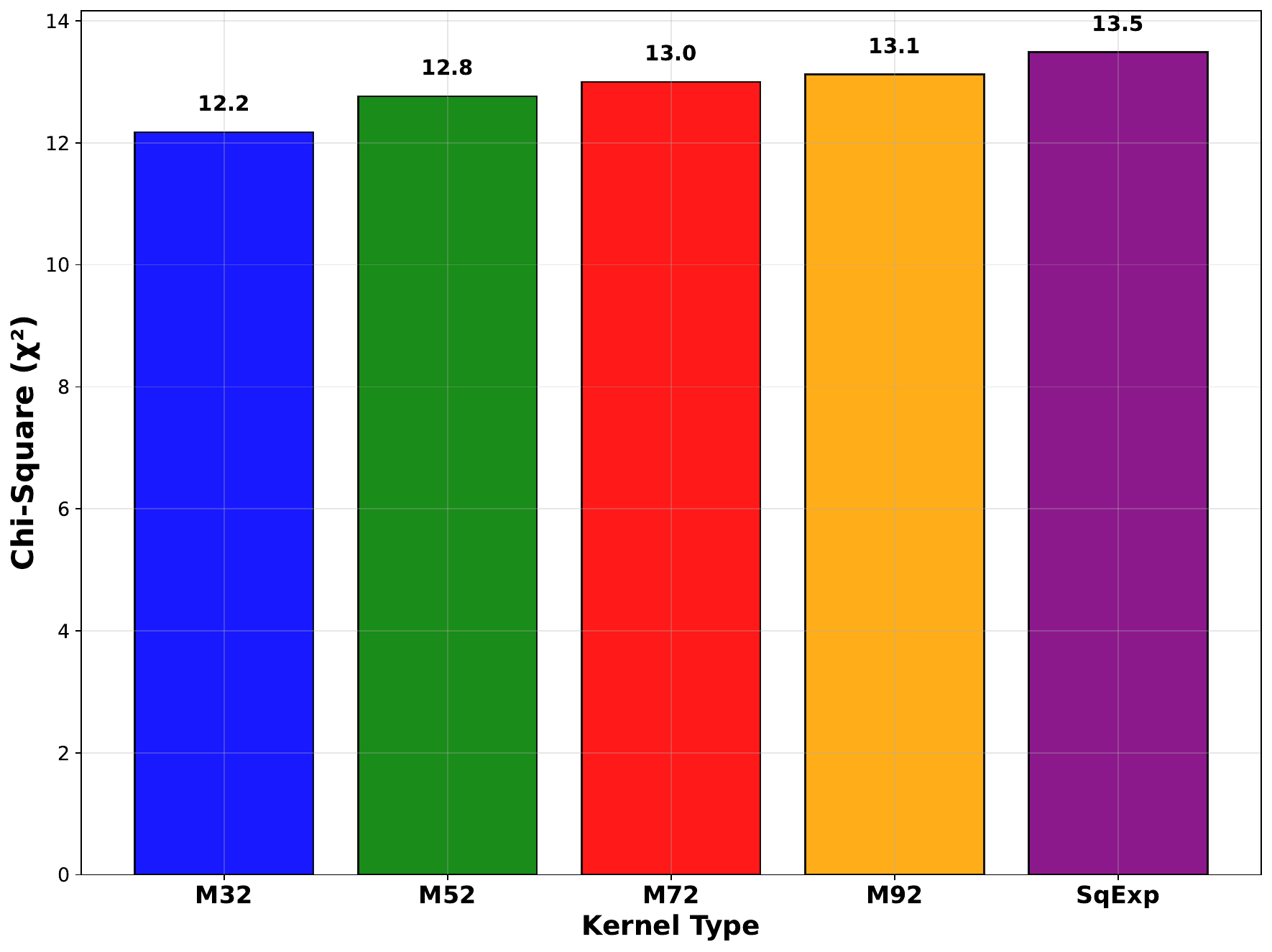}

\vspace{2pt}
{\small (a)}

\vspace{8pt}

\includegraphics[width=0.9\linewidth]{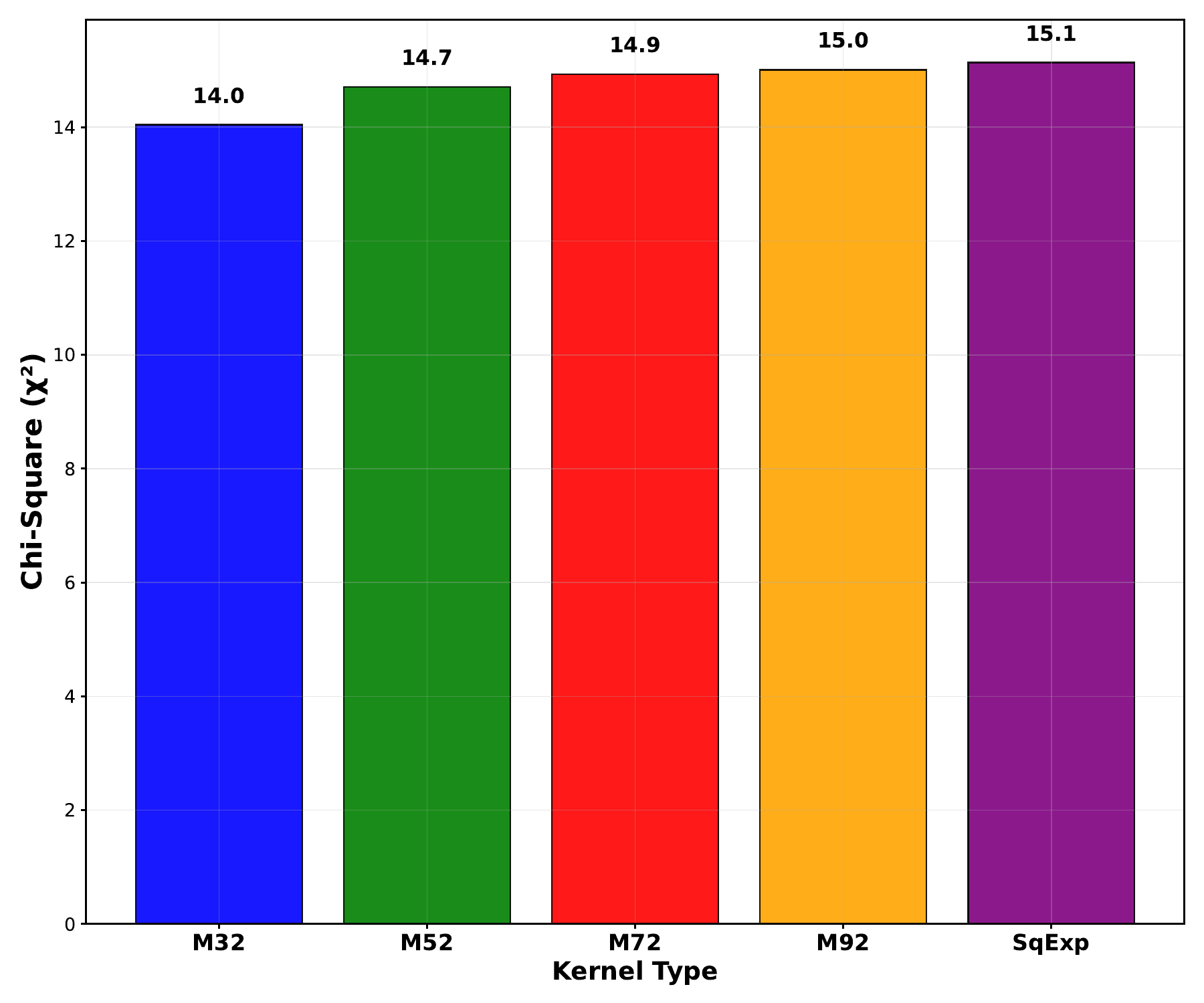}

\vspace{2pt}
{\small (b)}

\caption{
Chi-square values for different Gaussian Process kernels.
(a) Cosmic chronometer data (CC-32) only.
(b) Joint CC-32 + DESI BAO DR2 (6 DH/$r_{d}$) analysis.
A clear hierarchy $\chi^2_{3/2}<\chi^2_{5/2}<\chi^2_{7/2}<\chi^2_{9/2}<\chi^2_{\rm SE}$ is observed, demonstrating the systematic dependence of goodness of fit on kernel smoothness.
}

\label{fig:chi2}
\end{figure}

Further, we also examine whether the $\chi^2$ hierarchy among kernels persists when DESI BAO DR2 6 DH/$r_d$ data points\cite{DESI:2025zgx} are incorporated into the Gaussian Process reconstruction.
Even after jointly fitting CC and BAO observations, the ordering  
$\chi^2_{(3/2)} < \chi^2_{(5/2)} < \cdots < \chi^2_{\mathrm{SE}}$ remains clearly evident (Fig.~\ref{fig:chi2}b). This indicates that the relative smoothness of the kernel—not the specific dataset—drives the descriptive goodness-of-fit trend. The inclusion of BAO data tightens the cosmological constraints but does not alter the statistical hierarchy of the training residuals. Importantly, in light of the LML comparison discussed above, this ordering should not be interpreted as evidence that one kernel is favored by the data. Instead, it reflects how different smoothness priors redistribute residual structure while remaining statistically consistent within Bayesian evidence.

It is important to emphasize that the hierarchy observed in the
training-residual statistic $\chi^2$ (Fig.~\ref{fig:chi2}(a),(b)) characterizes the
in-sample flexibility of different kernels and should not be
interpreted as the direct cause of the hierarchy observed in
the reconstructed values of $H_0$ Fig.~\ref{fig:trend}. The two quantities
probe different aspects of the reconstruction: the $\chi^2$
ordering reflects how closely a kernel follows local features
of the data, while the $H_0$ ordering arises from how the
kernel smoothness and differentiability structure propagate
into the extrapolation toward $z=0$. The near-degeneracy of the
log marginal likelihood across kernels indicates that the data
do not strongly prefer one kernel over another, and therefore
the $\chi^2$ hierarchy should be interpreted descriptively
rather than as an explanation of the $H_0$ ordering.

As a robustness check, we tested our results through a jackknife resampling of the CC dataset. Removing individual data points does not alter the kernel hierarchy or the inferred trend in $H_0$, confirming that our conclusions are not sensitive to isolated measurements.

To provide an additional conservative test, we introduce a systematic floor of $5\,\mathrm{km\,s^{-1}\,Mpc^{-1}}$ added in quadrature to all reported CC uncertainties and repeat the GP reconstruction across the full kernel hierarchy. We find that the inferred value of $H_0$ shifts by at most $\sim 1\,\mathrm{km\,s^{-1}\,Mpc^{-1}}$, remaining well within the corresponding $1\sigma$ uncertainties, while the monotonic ordering with kernel smoothness is fully preserved. The optimized hyperparameters vary only at the few-percent level, confirming that our results remain stable even under deliberately inflated uncertainty assumptions.

\section{Discussion and conclusion}
In this work, we have investigated how the choice of covariance kernel in Gaussian Process regression influences the reconstruction of the Hubble parameter from cosmic chronometer data. Focusing on CC-32 measurements \cite{Zhang_2014, Stern:2010cv,Moresco:2012jh,Moresco_2016,10.1093/mnras/stx301,10.1093/mnrasl/slv037} and a systematic hierarchy of kernels—from the Matérn $3/2$ to the infinitely differentiable Squared Exponential—we have demonstrated that kernel smoothness acts as an effective prior that directly shapes the inferred cosmic expansion history. In particular, we find a clear and monotonic trend in the reconstructed Hubble constant: rougher kernels yield higher values of $H_0$, while progressively smoother kernels drive $H_0$ toward lower values with reduced uncertainties.

This kernel-dependent behavior provides a continuous, non-parametric interpolation between SH0ES-like \cite{Riess:2021jrx} and Planck-like\cite{Planck:2018vyg} determinations of $H_0$, achieved without invoking any specific cosmological model or dark energy parameterization. The result highlights that Gaussian Process reconstructions are not purely data-driven; rather, they encode strong assumptions through the kernel, especially regarding differentiability and correlation length. The commonly used Squared Exponential kernel, while mathematically convenient, appears comparatively restrictive for CC data under the adopted smoothness prior, suppressing local structure and leading to systematically lower inferred $H_0$ values.

Our $\chi^2$ analysis further supports this conclusion. We find a consistent hierarchy in goodness of fit,
$\chi^2_{(3/2)} < \chi^2_{(5/2)} < \chi^2_{(7/2)} < \chi^2_{(9/2)} < \chi^2_{\rm SE}$ (Fig.~\ref{fig:chi2}(a),(b)), reflecting the increasing rigidity of smoother kernels and their tendency to smooth over local fluctuations present in the data. Importantly, this ordering persists even when the DESI BAO DR2 6 DH/rs  measurements are incorporated, indicating that the observed trend is driven primarily by the smoothness of the kernel rather than by the specific dataset used. However, as indicated by the near-degeneracy of the maximized log marginal likelihood and LOO-CV across kernels, this $\chi^2$ hierarchy does not translate into decisive Bayesian evidence for a uniquely preferred covariance structure. Further  jackknife resampling of the CC data confirms that the kernel hierarchy and the inferred $H_0$ sensitivity are not driven by individual measurements.

Overall, our findings demonstrate that kernel selection is a critical and non-trivial component of cosmological machine learning analyses. We emphasize that conclusions drawn from Gaussian Process reconstructions—particularly in the context of the Hubble tension—must be interpreted jointly with the assumed kernel prior. This work provides practical guidance for kernel choice in cosmological applications and establishes a clear framework for understanding how derivative hierarchy and smoothness assumptions propagate into physical inference.

We stress that these kernel-dependent differences should be interpreted as prior sensitivity rather than bias relative to an unknown ground truth, since the true cosmic expansion history cannot be observed directly. Our results therefore quantify how smoothness assumptions propagate into cosmological inference, rather than identifying a uniquely correct kernel.

\section{Acknowledgment}
The author gratefully acknowledges Md Wali Hossain and Purba Mukherjee for valuable discussions and insightful suggestions that contributed to this work.
\bibliographystyle{elsarticle-num}
\bibliography{references}

\end{document}